\documentclass[shortnote]{jpsj3}

\usepackage{txfonts}
\usepackage{color}

\newcommand{\figwidth}{3.375in} 

\title{Computing a Knot Invariant as a Constraint Satisfaction Problem}

\author{Chihiro H. Nakajima\thanks{E-mail address: nakajima@stat.phys.kyushu-u.ac.jp}$^{1}$, and Takahiro Sakaue\thanks{E-mail address: sakaue@phys.kyushu-u.ac.jp}$^{1,2}$}

\inst{$^{1}$\address{Department of Physics, Kyushu University, 33 Fukuoka 812-8581, Japan}

$^{2}$\address{PRESTO, JST, Kawaguchi 332-0012, Japan}
}

\kword{statistical mechanics , Monte-Carlo simulation , knot invariant , constraint satisfaction problem , polymer knot}

\begin{document}
\maketitle

Knots are closed space curves embedded in three-dimensional (3-d) space. Classification of knot types depending on their topology is one of central issues in mathematical knot theory\cite{KK}. Several invariances, often in the form of polynomial functions, can be used for that purpose, calculation of which, however, is generally not an easy task. The computational cost $h$ of the knot invariances rapidly increases with the number $n$ of the crossing in the knot diagram (Fig. 1 and 2), e.g, at least $h \sim n^3$ for Alexander polynomial, and $h$ increases faster than any polynomial, i.e., NP(non polynomial)-problem for Jones polynomial\cite{DJAW}.

The $p$-colorability problem of knots provides a particular type of the invariance (see below for its definition).

There seems to be no generic method to find its \textit{solution}, i.e., a colored configuration with satisfied manner, called \textit{a coloring class}, and to count the total number of them.
In this note, we present a statistical mechanical formulation of the $p$-colorability problem of knots, which provides an algorithm to find the solution. The method also allows one to get some deeper insight into the complexity of the problem from the viewpoint of the constraint satisfaction problem.

\begin{figure}
\begin{center}
\includegraphics[width=0.38\textwidth]
{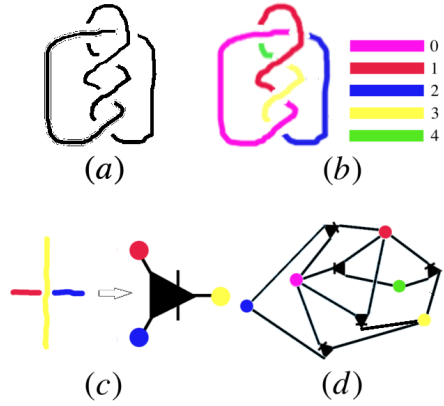}
\label{fig:proj_knot_diag_1}
\caption{(color online)(a):A schematic example of the knot diagram of figure-8 knot with $n=5$. (b):A satisfied colored pattern (coloring class) of the knot diagram. (c):A graphical reprsentation of the local satisfiability condition (constraint) on each crossing. A crossing and an arc are represented as a clause node (triangle) and a variable node (circle), respectively. Each clause is  connected with three involved variable nodes, where a node 
corresponding to the overcrossing arc is distinguished from other two. (d):A random graph corresponding to the knot (b). The number of edges connected with each variable node is random.}
\end{center}
\end{figure}

\textit{Formulation of the problem.} The knot topology can be analyzed by using the knot trajectory projected onto the plane with conserving over- and under- crossing conditions of local components.
It is called the knot diagram. As symbolically exemplified in Fig. 1(a), it consists of $n$ arcs and $n$ crossing points.
In $p$-colorability problem, we attempt to color each arcs using $p$ colors under a particular type of the constraint (see Fig. 1(b) ).
Let $c_i \in \{0 \ , \ \cdots \ , \ p-1\}$ be the color of $i-$th arc.
Each crossing point consists of three arcs, at which a constraint is defined locally.
Suppose that two arcs $i$ and $i+1$ are separated by an overcrossing of an arc $k$ at $i-$th cross. 
If the following equality holds among the colors of these arcs $c_i$, $c_{i+1}$ and $c_k$,
\begin{eqnarray}
\mathrm{mod}(c_i+c_{i+1},p) = \mathrm{mod}(2c_k,p),
\end{eqnarray}
the $i-$th cross is satisfied.
The total number $N(p;\mathcal{K})$ of coloring classes which satisfy all the constraints imposed at each crossings is known as one of the knot invariants, where $\mathcal{K}$ stands for the knot type.
The inequality $N(p;\mathcal{K}) \neq N(p;\mathcal{K}')$ guarantees the different topology $\mathcal{K} \neq \mathcal{K}'$, thus, $N(p;\mathcal{K})$ is the knot invariant.
However, like any other invariants, the reverse is not generally valid, \textit{i.e.}, the value of $N(p;\mathcal{K})$ and $N(p;\mathcal{K}')$ are not always different for different knot types.

\textit{Mapping onto constraint satisfaction problems.} By introducing two types of nodes, we obtain a graphical representation of a knot diagram (see Fig. 1(d) ).
Variable nodes (circles of Fig.1(c) and (d)) and clause nodes (triangles) correspond to  arcs and crossing points, respectively, which have the following connection properties,
\begin{description}
\item[(\textit{i})]Associated with each clause node are three edges connected to variable nodes; within $i$-th clause node, $i$-th , $i+1$-th , and $k$-th variable nodes are involved (Fig.1(c)).
\item[(\textit{ii})] The number $m(i)$ of edges connected to $i$-th variable node is random under the constraint $\sum_{i=1}^{n}m(i)=3n$.
\end{description}
Hence, graphs corresponding to knots have equal number $n$ of clause and variable nodes, and are allowed to have random connecting property patterns (by (\textit{ii})) with the condition (\textit{i}) imposed.
The satisfiability function $h_i(c_i,c_{i+1},c_k)$is defined as,
\begin{eqnarray}
h_i(c_i,c_{i+1},c_k)=\left\{
\begin{array}{l}
1 \quad \big( \ \mathrm{mod}(c_i+c_{i+1},p)-\mathrm{mod}(2c_k,p)\ne0 \ \big) \\
0 \quad \big( \ \mathrm{mod}(c_i+c_{i+1},p)-\mathrm{mod}(2c_k,p)=0 \ \big)
\end{array}
\right. .
\end{eqnarray}
The degree of the net satisfiability is quantified by the function
\begin{eqnarray}\label{eq:Hamiltonian_basic}
H=\sum_{i=1}^{n}h_i(c_i,c_{i+1},c_k).
\end{eqnarray}

Thus a projected knot diagram is mapped onto a random graph, and the problem of computing $N(p;\mathcal{K})$ is mapped onto the problem of counting the number of ground states of the Hamiltonian (\ref{eq:Hamiltonian_basic}); a constraint satisfaction problem on random graph. 

\textit{Inplement of simulation.} We performed replica exchange Monte Carlo simulations\cite{HN} of the Hamiltonian (\ref{eq:Hamiltonian_basic}) and estimated the ground state entropy, which corresponds to the logarithm of $N(p;\mathcal{K})$.
For computation of the entropy $S(\beta)$ at each inverse temperature point $\beta=1/T$, we used following thermodynamic integral equation,
\begin{eqnarray}
S(\beta)=S(0)+\int_0^{\beta}U(\beta')d\beta',
\end{eqnarray}
where $U(\beta)$ represents the internal energy.

\begin{figure}[!htbp]
\includegraphics[width=\figwidth]
{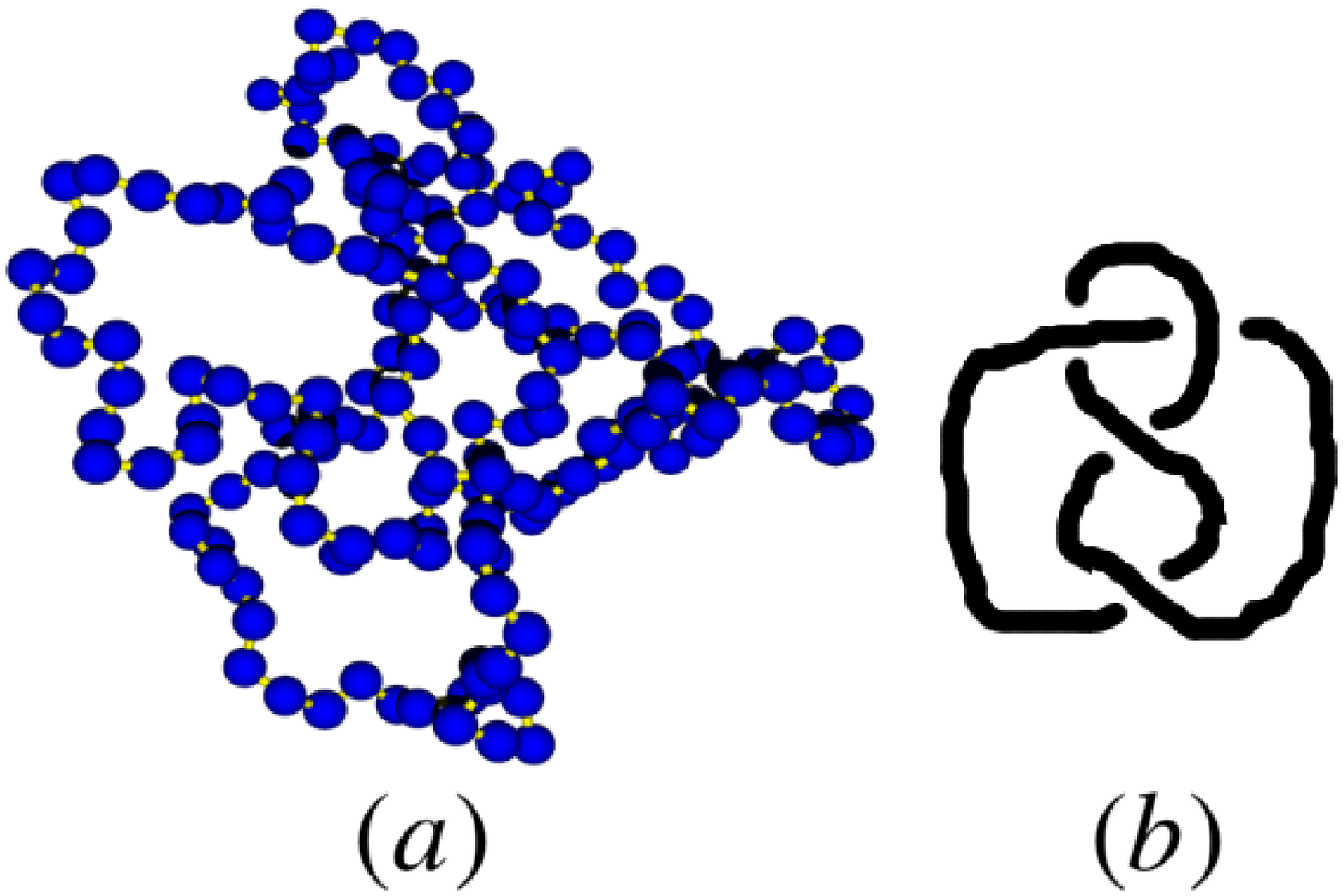}
\caption{(color online)
(a):Conformation of figure-8 knot used to perform trial computation. The corresponding knot diagram has 26 crossings. (b):A realization of the minimal crossing points; $4$ of figure-8 knot.
The diagram(a) with $n=26$ and that in Fig.1 with $n=5$ can be transformed to the diagram (b) with $n=n_{min}=4$ via Reidemeister moves.}
\label{fig:conformation}
\includegraphics[width=\figwidth]
{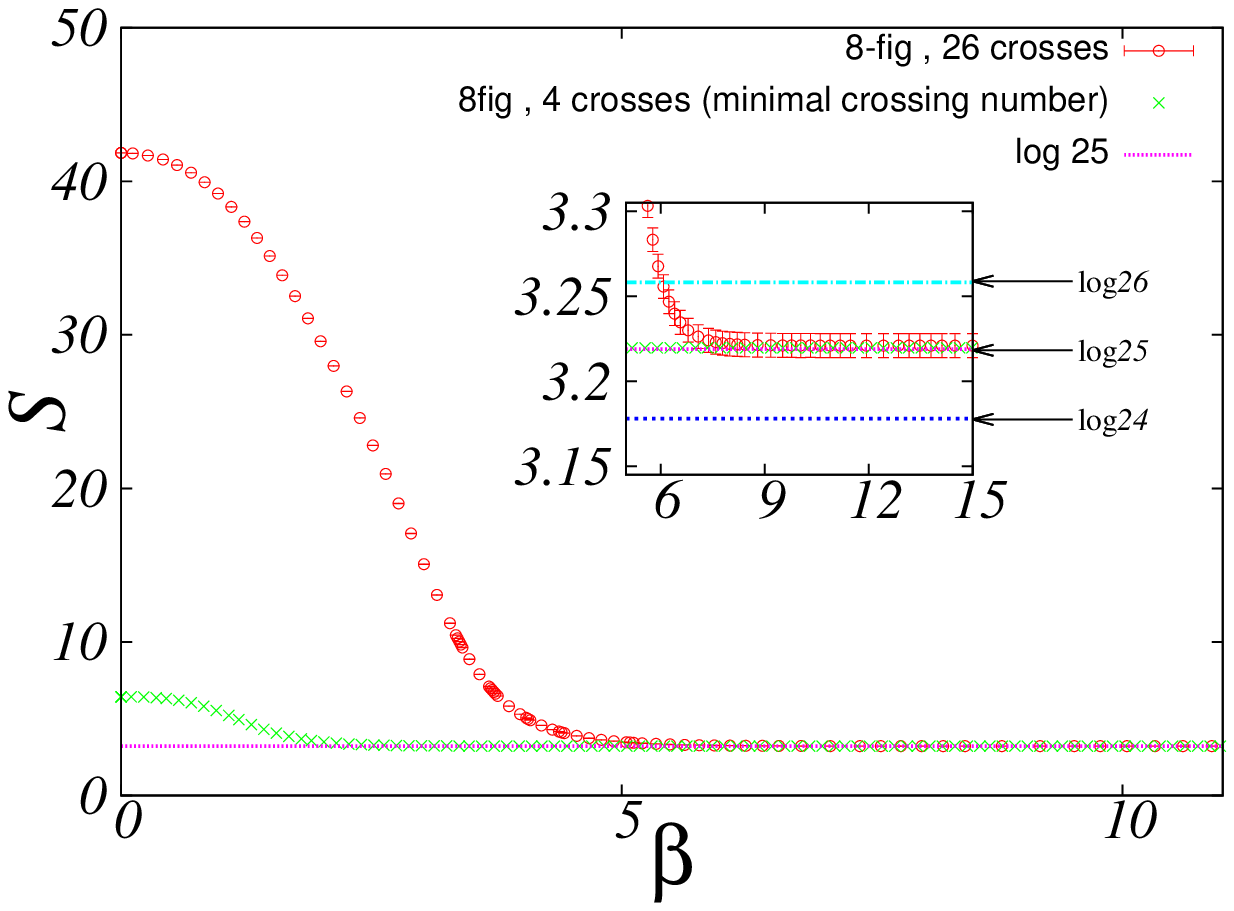}
\caption{(color online)Computed entropy profile for $n=26$ system (Fig.\ref{fig:conformation}(a)) as a function of inverse temperature (circles). Error bars result from the average of 10 independent simulations. For comparison the result of minimal crossing number ($n=4$) system (Fig.\ref{fig:conformation}(b)) is also shown (crosses), which was obtained by exact enumeration of Hamiltonian (\ref{eq:Hamiltonian_basic}). At low temperature, both converge to the value $\log 25$ within the accuracy of the numerical error, see the close up (inset).
}
\label{fig:entropy_profile}
\end{figure}

First we generated  3-d random sample trajectories with figure-8 knot type by performing a standard Langevin-dynamics (LD) simulation using a closed beads-spring model polymers (Fig.\ref{fig:conformation}). Random graphs were obtained by projecting the trajectories onto the plane.
We then carried out MC simulation explained above for each graph and obtained entropy as a function of inverse temperature (Fig.\ref{fig:entropy_profile}).
To obtain smooth profiles in Fig.\ref{fig:entropy_profile}, simulations were performed on 40 different temperatures and gaps are interpolated using histogram reweighting technique.
The number of coloring classes evaluated as an exponentiation of the ground state entropy agrees with the known value, 
i.e., 25 for a figure-8 knot with $p=5$ colors.
At this example, the number of coloring class with satisfied manner is prosperously evaluated. 

\textit{Discussion and perspectives} The present method is, in principle, applicable to much more complex knots than considered here with larger minimal crossing numbers.
The results from several different coloring number $p$, and if necessary combined 
with other kind of the invariants, would have a good classification ability of knots.

So far we have only focused on the ground states of the model Hamiltonian eq.(\ref{eq:Hamiltonian_basic}).
Ground states are obviously most important in the context of knot theory with a definite meaning corresponding to the invariance. It does not, however, necessarily exclude the possible implications of the {\it excited states}.
For instance, we expect that the internal energy of our model at finite temperature may contain some useful information on the knot complexity.

The coloring problem on random graphs often show characteristic phase transition behavior involving both statics\cite{ZK2007} and dynamics\cite{S2008} from simple structureless phase to glassy phase with many metastable states\cite{MPWZ,KMRTSZ}.
It has been intensively studied from the spin glass perspective\cite{MM} and often discussed in relation to computational complexity\cite{MPWZ,MZKST}.
A unique feature in the present system comes from the fact that the ground states of our model are connected to the topological invariance. Therefore, we can control the apparent conformational complexity in arbitrary ways, i.e., the crossing number $n$ in the knot diagram, while keeping the ground states invariant via Reidemeister moves\cite{KK} (See the caption in Fig.\ref{fig:conformation}).
The model surely exhibits extensively separated ground states, but its total number does not grow exponentially with the system size $n$.
This fact means that the model does not exhibit clustering and condensation transitions on solution space structure\cite{KMRTSZ} in literal terms. 
Detailed investigations on such a model system may be interesting towards better understanding of the glassy properties with rugged landscapes\cite{KZ2010}
 and empirical hardness of searching problems\cite{MH,ZM,CMMS}.


\end{document}